# Anomalous Magnetoresistance by Breaking Ice Rule in $Bi_2Ir_2O_7/Dy_2Ti_2O_7$ Heterostructure

H. Zhang[1†], C. K. Xing[1†], K. Noordhoek[1†], Z. Liu[2], T. H. Zhao[3], L. Horák[4], Q. Huang[1]. L. Hao[1], J. Yang[1], S. Pandey[1], E. Dagotto[1,5], Z. Jiang[3], J. -H. Chu[2], Y. Xin[6], E. S. Choi[6], H. D. Zhou[1*] and J. Liu[1*]

[1]Department of Physics and Astronomy, University of Tennessee, Knoxville, TN, USA, 37996
[2]Department of Physics, University of Washington, Seattle, WA, USA, 98195
[3]School of Physics, Georgia Institute of Technology, Atlanta, GA, USA, 30332
[4]Charles University, Prague, Czechia,11636
[5]Materials Science and Technology Division, Oak Ridge National Laboratory, Oak Ridge, TN, USA, 37831
[6]National High Magnetic Field Laboratory, Florida State University, Tallahassee, FL, USA, 32310

†equal contribution; *corresponding author

While geometrically frustrated quantum magnets[1] are known for a variety of exotic spin states that are of great interests of understanding emergent phenomena as well as enabling revolutionary quantum technologies[2-4], most of them are necessarily good insulators which are difficult to be integrated with modern electrical circuit that relies on moving charge carriers. The grand challenge of converting fluctuations and excitations of frustrated moments into electronic responses is finding ways to introduce charge carriers that interact with the localized spins without destroying the spin states. Here, we show that, by designing a $Bi_2Ir_2O_7$/$Dy_2Ti_2O_7$ heterostructure, the breaking of the spin ice rule in insulating $Dy_2Ti_2O_7$[5-7] can lead to a charge response in the $Bi_2Ir_2O_7$ conducting layer that can be detected as anomalous magnetoresistance. These results demonstrate a novel and feasible interfacial approach for electronically probing exotic spin states in insulating magnets, laying out a blueprint for the *metallization* of frustrated quantum magnets.

Novel emergent states of quantum magnets often arise from geometric frustration, where local spins residing on the corners of a triangle or tetrahedron cannot agree on any alignment that simultaneously minimizes their energies[1]. Such a spin lattice has a large degree of degeneracy with an enormous number of configurations sharing the same or similar energy, leading to exotic ground states, entangled fluctuations, and collective excitations[8-13]. The so-called dipolar spin ice is one of the most iconic representatives of such frustrated magnets. It was first discovered on pyrochlore lattices of Ising spins constrained along the local [111] direction, such as $Ho_2Ti_2O_7$ and $Dy_2Ti_2O_7$[5-7]. While the spin-spin correlation is dominated by the dipolar interaction, the ferromagnetic instability is geometrically frustrated by the local Ising anisotropy[1]. As a result, the ground state of each tetrahedron settles in one of the six-fold degenerate 2-in-2-out configurations, forming a spin ice network throughout the lattice following the ice rule in analog with the proton displacement in the water ice shown in Fig. 1(a)[14]. The non-zero residual entropy has indeed been observed with a value close to that in the water ice[15]. Breaking the ice rule could lead to rich spin dynamics, such as emergent magnetic monopoles[16]. For instance, applying a magnetic field along the [111] direction turns the three-dimensional spin ice into a Kagome spin ice[17], and eventually stabilizes the 3-in-1-out/1-in-3-out configuration by condensing the monopoles, giving rise to a jump from ~3.3 $\mu_B$/Dy to ~5 $\mu_B$/Dy in $Dy_2Ti_2O_7$ between two magnetization plateaus[18]. The 3-in-1-out/1-in-3-out state is also called the saturated ice[19] where the ice rule is broken everywhere.

However, the spin-ice behaviors have been only observed in highly insulating compounds[1,5,8], where the spins are introduced by highly localized electrons in the absence of itinerant carriers, such as the *f*-electrons of $Ho^{3+}$ or $Dy^{3+}$ ions in combination with the empty *d*-shell of the $Ti^{4+}$ valence state in the pyrochlore titanates[5-7]. While it is an interesting open question whether and how the spin ice state would interact with itinerant carriers and introduce any novel electronic behaviors, simply replacing the B-site with another element with a metallic configuration often ends up destroying the spin ice state. For instance, the correlated *d*-electrons in pyrochlore iridates not only exhibit an all-in-all-out ordering of their own at much higher temperatures, but also force the same magnetic order onto the rare earth sublattice[20]. Although the rare earth moments may in turn modify the transport properties of the *d*-electrons[21], the physics is still dictated by the energy scale of the spontaneous magnetic order of the *d*-electrons, which is much larger than the interaction within the rare earth sublattice. This issue indeed applies to a broad class of frustrated magnets beyond spin ice.

In this work, we report the observation of an anomalous magnetoresistance (MR) in epitaxial pyrochlore heterostructures of $Bi_2Ir_2O_7/Dy_2Ti_2O_7$ (BIO/DTO), where DTO hosts the spin ice state and BIO, the only nonmagnetic/time-reversal-invariant member of the pyrochlore iridate series, provides the correlated carriers. We directly grow ultrathin BIO films on DTO single crystals as illustrated in Fig. 1(b). An MR anomaly is observed, when the field induces the transition between the Kagome spin ice state and the saturated ice. These results demonstrate that having an epitaxial interface is an effective approach to inducing charge carriers that not only preserve but also interact with the exotic spin states in insulating frustrated magnets.

DTO single crystals were prepared by the floating zone method[22]. The crystals were first oriented and cut into substrate pieces along the DTO (111) lattice plane. The substrate surfaces were then polished through a five-step process (see details in experimental method) to achieve atomically flat surfaces. Figure 1(c) shows a typical atomic force microscope (AFM) image of the DTO single crystal substrate surface after the process, with a surface roughness of ~1.21 Å. Thin BIO layers of 3 to 5 nm were deposited on the DTO substrates by pulsed laser deposition[23]. Figure 1(d) shows a typical specular X-ray diffraction scan covering the DTO [222] to [444] structural peaks, which are accompanied with the corresponding peaks of the BIO film, confirming the epitaxial growth. Reciprocal space mapping (RSM) shows that the BIO film is in a fully strained state (Figure S1 in the supplementary). The thickness and density of the film extracted from the X-ray reflectivity curve (Figure S2) are consistent with expectations. The BIO/DTO heterostructure was also examined with the scanning transmission electron microscopy (STEM). A typical cross-section image in Fig. 1(e) shows a sharp epitaxial interface. Energy dispersive X-ray spectroscopic (EDS) map (Figure S3) further confirms no significant interdiffusion across the interface.

While the all-in-all-out magnetic ordering of the Ir pseudospin-1/2 electrons in the pyrochlore iridates has an increasing ordering temperature with decreasing A-site ionic radius[24], BIO is a bad metal on the paramagnetic side of the quantum critical point[25]. No magnetic ordering or spontaneous time-reversal-symmetry-breaking was found down to 50 mK[26]. Moreover, it has been shown that, although the BIO metallicity will degrade when grown as ultrathin films likely due to weak localization and/or disorder-enhanced electron–electron interaction in two-dimensions, the MR remains rather isotropic when the field is applied along different crystallographic axes[27]. Our ultrathin BIO films on DTO indeed show a weakly increasing resistivity as lowering temperature as well as an isotropic MR (Figure S4 in the supplementary), consistent with previous reports[27]. However, an emergent anisotropic anomaly was observed while measuring MR below 1 K. Figure 2(a-b) shows MR measurements at 0.03 K with the magnetic field $B$ along different directions in the (11-2) plane, while the current is always along the [11-2] axis perpendicular to the field. One can see the overall MR response with $B$//[111] is similar to $B$//[1-10] except a positive anomaly around 1.5 T. Our AC susceptibility measurement shows that this field corresponds to the critical field $B_c$ of the Kagome spin ice-to-saturated ice transition of DTO, and it is higher than the intrinsic value of $B_c$ ~ 1 T[19] due to the demagnetization effect of the plate-shape of the crystal substrate (see Fig.1(b) & Fig. S5). Moreover, as we rotate $B$ away from [111], the anomaly displays a clear shift to higher fields with the width of the feature being broadened as well, which is consistent with the fact that the field projection to [111] is decreasing as the field angle increases. A larger external field is thus needed to induce the transition, and the finite width of the transition is effectively stretched. The anomaly is eventually shifted beyond the measured field range, by which point the MR curve is virtually the same as that with $B$//[1-10]. We have verified these observations by measuring a reference sample where the BIO film was deposited on a (111)-oriented $Y_2Ti_2O_7$ (YTO) single crystal substrate, and the MR indeed is isotropic between $B$//[111] and $B$//[1-10] without any extra feature (Figure S6 in the supplementary), consistent with the fact that Dy is replaced with the nonmagnetic Y ion. In summary, this emergent MR anomaly of the BIO film is clearly associated with the magnetic response of DTO.

Figure 2(c-d) shows representative MR curves of the BIO/DTO heterostructure at 0.03 K measured in another configuration where $B$ is applied in the (1-10) plane with current along the [1-10] axis. This setup allows MR measurement with the field along multiple high-symmetry axes, including [110], [11-1], and [001], in addition to [111] as shown in Fig. 2(c). One can see that, while the overall MR response remains similar for all directions, the MR curve with $B$//[11-1] clearly shows an extra positive anomaly that is absent with $B$//[110] and $B$//[001] but is similar to the case $B$//[111]. This observation is consistent with the fact that [111] and [11-1] are two equivalent axes in DTO. The difference is that the anomaly with $B$//[11-1] is at ~1 T, which is lower than that with $B$//[111] and close to the intrinsic $B_c$. We again confirmed this behavior by AC susceptibility measurements, which show in the bottom panel of Fig. 2(d) that the peaks in correspondence to the transition for $B$//[111] and $B$//[11-1] indeed occur at different fields and match the fields of their MR anomalies, respectively. The shift of $B_c$ between [111] and [11-1] is attributed to the much stronger demagnetization effect for the surface normal direction [111]. By comparison, the AC susceptibility is featureless for $B$//[110] and $B$//[001]. All the observations above corroborate that the MR anomaly of the BIO layer is indeed a result of the field-induced ice-rule-breaking transition in DTO and is remarkably sensitive to any shift of $B_c$.

This result demonstrates that the interfacial approach can effectively introduce interactions between the charge carriers from one side and the frustrated spins on the other side. Note that the field, regardless of its direction, always locally flips the moments to partially lift the 2-in-2-out degeneracy of the Dy tetrahedron within the spin ice state. Yet the MR anomaly only occurs when the transition to the saturated ice state is induced, indicating that the interfacial interaction renders the charge carriers particularly sensitive to the ice-rule-breaking spin flips. Given that the Kagome spin ice-to-saturated ice transition is a first order transition[1,19,20], the most likely mechanism of the anomalous MR responses is the coexistence of the two spin states in the Kagome plane perpendicular to the field (since the triangular plane is polarized in both cases). As illustrated in Fig. 3(a), a boundary of the two regions is defined by the spins that connect tetrahedra hosting the 2-in-2-out and 3-in-1-out/1-in-3-out configuration (Fig. 3(b)), respectively.

The spins nearby the boundary are most unstable and thus act as scattering centers for the charge carriers when entering one region from another. Also illustrated in Fig. 3(a) is a cartoon of how the boundary density may increase and decrease in the critical regime where the regions of the two states repopulate. This picture captures the positive sign of the anomaly. It is also consistent with the absence of any significant vertical shift or jump of the MR curve before and after the transition. In other words, the anomaly is a charge response to the magnetic fluctuations in the transition region, and it has a peak-like line shape similar to the AC susceptibility. We note that the spin Hall effect (SHE) in metals may also produce anisotropic MR responses when in direct contact with an insulating magnet[28]. This so-called spin hall magnetoresistance however requires an orthonormal geometry among the surface normal, current direction and magnetization, which is obviously unnecessary for the observed anomaly in the BIO/DTO heterostructure, excluding the SHE for being the underlying mechanism.

Studies of the field-temperature phase diagram of spin ice have shown that the first order phase boundary is terminated at elevated temperature and the transition becomes a continuous crossover above the freezing temperature as there is no sharp distinction between the Kagome spin ice and the saturated ice anymore[8,18]. This behavior has been understood in analogy with the liquid–gas transition in the context of a transition between a low-density phase and a higher-density phase of magnetic monopoles [16-18]. For comparison, we studied the thermal evolution of the MR anomaly with *B*//[111] between 0.45 K and 1 K. The obtained result displayed as the solid lines in Fig. 3(c) shows that, while the anomaly overall remains around 1.5 T, it weakens in magnitude, broadens in width, and upshifts slightly in field with increasing temperature, likely due to the thermal broadening of the Kagome spin ice-to-saturated ice transition[1]. The dashed lines were measured with *B*//[1-10] and serve as a reference for comparison at each temperature. The anomaly eventually starts to vanish at ~0.9 K and becomes hardly visible. These observations are consistent with the field-temperature phase diagram as well as the melting of the spin ice state in DTO due to thermal fluctuation[14-16], confirming the sensitivity of the charge carriers to the abrupt change between the two spin states until they are no longer distinguishable above the critical endpoint of the transition[18].

To summarize, we experimentally demonstrated that having an epitaxial interface is a highly promising approach to introducing interactions between localized frustrated spins and correlated charge carriers, by observing an emergent MR anomaly in BIO in response to the breaking of the ice rule in insulating DTO and closely tracking the transition to the saturated ice state. It opens the possibilities of electronically probing the spin ice states under other control parameters, such as pressure and epitaxial strain[29], as well as investigating charge conversion of the monopole[16] excitations under non-equilibrium stimuli[30,31]. The interfacial approach can readily be extended to other non-conducting frustrated quantum magnets, such as spin liquid and quantum spin ice[1-4,9-13], for electronically detecting these exotic spin states and obtaining novel electronic phenomena. The successful creation of functional epitaxial interfaces between DTO and BIO also showcases the potential of complex oxide heterostructures of the pyrochlore structure[32].

**Acknowledgement:** This research is supported by the U. S. Department of Energy under grant No. DE-SC0020254. H. D. Z. and J. L. acknowledge support from the Organized Research Unit Program; the Electromagnetic Property Laboratory; the Scholarly Activity and Research Incentive Fund (SARIF) at the University of Tennessee. E. D. was supported by the US Department of Energy (DOE), Office of Science, Basic Energy Sciences (BES), Materials Sciences and Engineering Division. The work at University of Washington is supported by the David and Lucile Packard Foundation (Transport measurements) and the Air Force Office of Scientific Research through DURIP Award FA9550-20-1-0310 (Helium 3 cryostat). The transport measurement at Georgia Tech is supported by the DOE under grant No. DE-FG02-07ER46451. The synchrotron X-ray diffraction measurement used resources of the Advanced Photon Source, a U.S. Department of Energy (DOE) Office of Science User Facility operated for the DOE Office of Science by Argonne National Laboratory under Contract No. DE-AC02-06CH11357. A portion of this work was performed at the National High Magnetic Field Laboratory, which is supported by the National Science Foundation Cooperative Agreement No. DMR-1644779 and the state of Florida. The authors thank Martin Mourigal and his research group for providing help in resistivity measurements; Jenia Karapetrova for helping with the synchrotron X-ray diffraction; and Cristian Batista for fruitful discussions.

**Experimental Method:**

**Synthesis of the DTO single crystal:** polycrystalline DTO was prepared by thoroughly mixing stoichiometric $Dy_2O_3$ and $TiO_2$ with the molar ratio of 1:2. Prior to weighing, $Dy_2O_3$ was preheated at 1000 degree for 10 h in order to remove the moisture. After grinding the mixture, sintering the mixture at 1000 degree and 1400 degree for 20 h each in air with intermediate grinding. The powders were then compressed to a feed rod with another additional annealing at 1400 degree for 20 h, which were used to grow the single crystal DTO by using a double elliptical mirror optical floating-zone furnace SC1-MDH from Canon Machinery. To avoid oxygen deficiency, we grew the single crystal in 5 atm flowing $O_2$ atmosphere. In the first cycle we used a high growth rate of 50 mm/h for both feed and seed rods and in the second cycle we used a relatively slow growth rate of 4 mm/h.

**Sample preparation:** The DTO single crystal was first aligned by a HUBER Laue X-ray diffractometer and then cut into substrate pieces with DTO (111) lattice plane by using the Model 650 Low Speed Diamond Wheel Saw from South Bay Technology. The substrate surfaces were prepared by going through a five-step mechanical polishing process with monocrystalline suspension. The surfaces were characterized by an MFP-3D Atomic Force Microscope from Asylum Research. Thin film samples were synthesized by Pulsed Laser Deposition[23]. A KrF excimer laser ($\lambda$ = 248 nm) was used at a repetition rate of 2 Hz and the laser beam fluency is 1.7 J/cm$^2$. The BIO films were deposited at 620 degree and 0.0667 mbar of oxygen pressure. The thickness was controlled by the calibrated growth rate of BIO films at 40 pulses/nm. After deposition, the films were cooled down in 0.0667 mbar oxygen pressure.

**X-ray characterization:** In-house X-ray characterization was done by an Empyrean diffractometer at Copper radiation. Synchrotron X-ray diffraction was performed at beamline 33-BM at the Advanced Photon Source at the Argonne National Laboratory.

**Transmission Electron Microscopes (TEM):** TEM data were acquired using a cold field emission probe-aberration-corrected JEOL JEM-ARM200cF at 200kV with a spatial resolution of 0.078 nm. It is equipped with Oxford Aztec SDD EDS detector for elemental analysis. A thicker film sample with thickness ~10.5nm was used in the TEM study to protect the interface. TEM lamella was made by focused ion beam (FIB) in a ThermoFisher Scientific Helios G4 UC DualBeam scanning electron microscope (SEM). To avoid charging in the SEM, the thin film surface was first sputtered with a 10 nm thick layer of gold. Cross-sectional view along [1-10] was obtained by high-angle annular dark-field scanning transmission electron microscopy (HAADF-STEM) imaging. The HAADF-STEM images were collected with the

JEOL HAADF detector using the following experimental conditions: probe size 7c, CL aperture 30µm, scan speed 32 µs/pixel, and camera length 8 cm, which corresponds to a probe convergence angle of 21 mrad and inner collection angle of 74 mrad. EDS maps were collected in the STEM mode with a probe size of 0.12 nm.

**Resistivity and magnetoresistance measurements:** Transport measurement was performed with a Physical Properties Measurement System (PPMS) from Quantum Design (QD). The current is always applied perpendicular to the field direction. MR between 0.45 K and 1 K was measured with the He-3 option of a PPMS Dynacool from QD. MR at 0.03 K was measured at the National High Magnetic Field Laboratory. All MR data were measured by utilizing the lock-in technique. Samples used in the measurement are of 3 mm (W) × 6 mm (L) in dimension. The DTO single crystal is of 0.5 mm in thickness.

**AC susceptibility:** AC susceptibility was measured at the National High Magnetic Field Laboratory. The AC susceptibility measurements were conducted with a voltage controlled current source (Stanford Research, CS580) and lock-in amplifier (Stanford Research, SR830). The phase of the lock-in amplifier is set to measure the first harmonic signal. The rms amplitude of the AC excitation field is set to be 0.35 Oe with the frequency fixed to be 470 Hz.

**Author contributions:** H. D. Z and J. L. conceived the idea and directed the study. K. N. and C. K. X. performed the pulsed laser deposition. C. K. X. synthesized the single crystal substrates. K. N. and C. K. X. performed the surface preparation. L. H. and C. K. X. performed in-house X-ray diffraction measurement. H. Z. performed synchrotron X-ray diffraction with the help of L. H. and J. Y.. H. Z., Z. L., J. H. C., T. H. Z, Z. J., C. K. X., K. N., S. P. and E.S.C. performed the transport measurements. C. K. X. and Q. H. conducted the ac susceptibility measurement. H. Z., K. N., C. K. X., H. D. Z, and J. L. analyzed the data. H. Z., H. D. Z, E. D. and J. L. wrote the manuscript with contributions from all other authors.

**Competing interests**

The authors declare no competing interests.

**Figure 1**

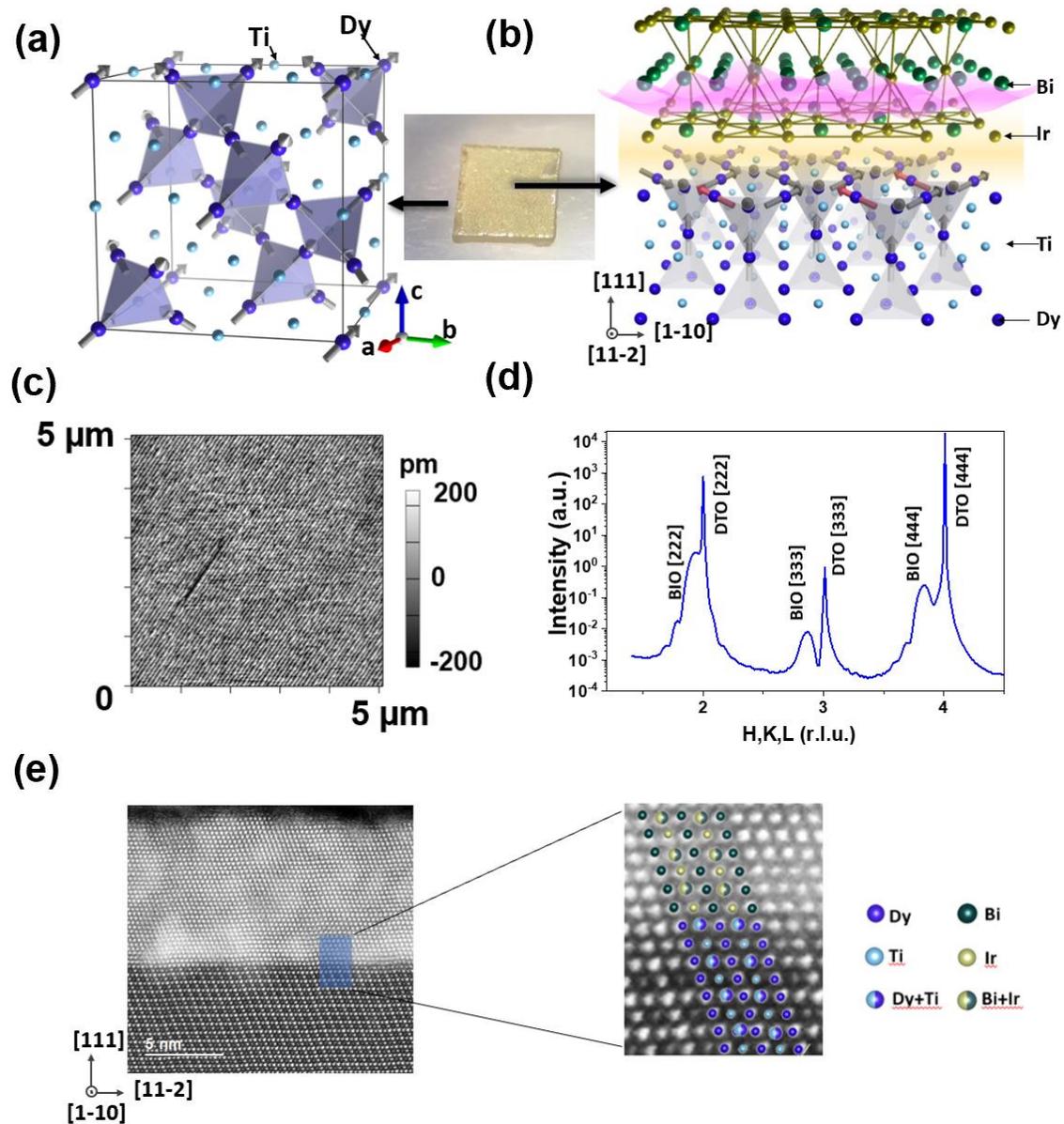

Figure 1 (a) Dy$_2$Ti$_2$O$_7$ crystal and spin structure, the Dy and Ti atoms each form corner-sharing tetrahedra. Oxygen atoms are not shown for clarity purpose. The spins on the Dy sites show the '2-in-2-out' configuration. (b) A picture of a (111)-oriented DTO crystal where a BIO film is deposited on the top to form the pyrochlore heterostructure as illustrated in the schematic drawing on the right. The local spins (grey) are shown on the DTO sides with a few of them (red) flipped to break the ice rule and form the '3-in-1-out/1-in-3-out' configuration. A wavy surface curve (pink) is shown on the BIO side to denote charge carrier. (c) A typical AFM image of DTO substrate surface. (d) An X-ray diffraction scan covering the DTO [222] to [444] peaks of a BIO/DTO heterostructure. (e) Atomic resolution high-angle-annular-dark field-STEM image of the interface. The TEM lamella surface normal is [1-10]. A zoomed picture across the interface is shown on the right to illustrate the sharp interface of the heterostructure.

**Figure 2**

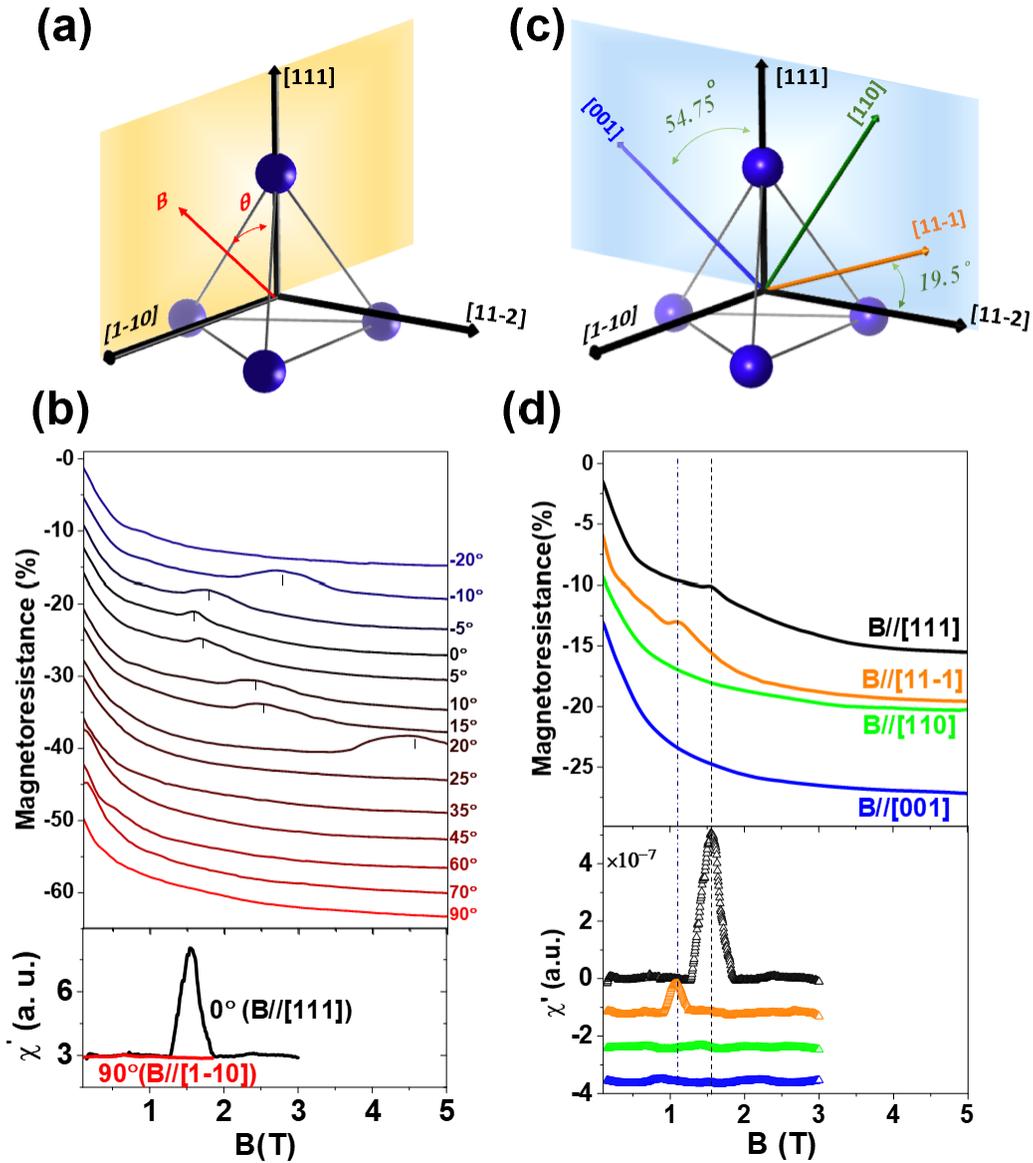

Figure 2 (a) Schematic diagram of the MR measurement with the field in the (11-2) plane, where θ is the field angle respect to the [111] axis. (b) Measured MR (top) and AC susceptibility (bottom) for the BIO/DTO heterostructure at 0.03 K at selected angles. Vertical offsets are applied on the curves for clarity purpose. (c) Schematic diagram of the MR measurement with the field in the (1-10) plane, which contains four high-symmetry axes as indicated. (d) Measured MR (top) and AC susceptibility (bottom) at 0.03 K along the high-symmetry axes in this geometry. Vertical offsets are applied on the curves for clarity purpose.

**Figure 3**

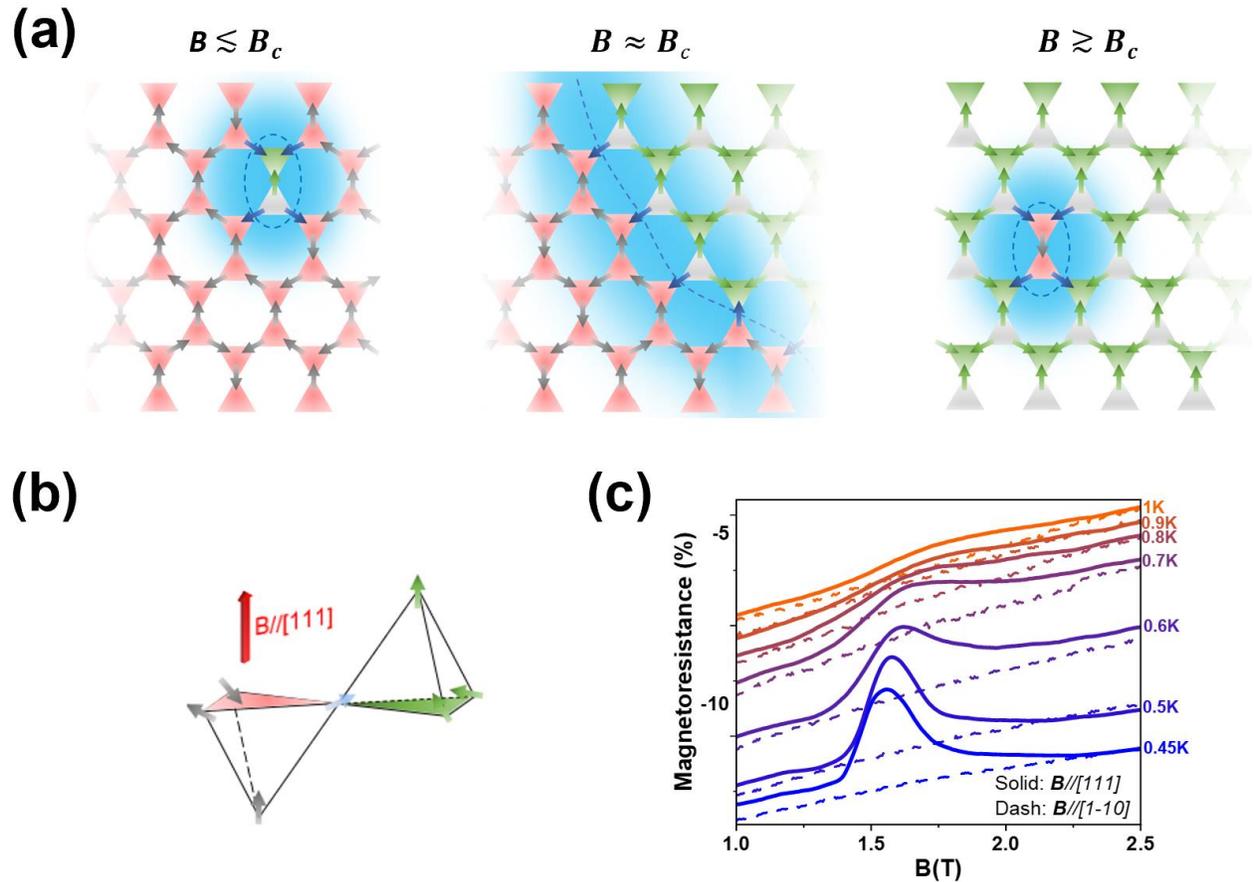

Figure 3 (a) Schematic diagram showing the boundary in the Kagome layer between the Kagome spin ice and the saturated ice state. The red triangles represent the tetrahedra that have the 2-in-2-out configuration, whereas the green (grey) ones indicate those in the 3-in-1-out (1-in-3-out) configuration. The blue arrows denote the sites that connect the two regions with the area highlighted in light blue indicating the boundary region where the spins are less stable. A three-dimensional schematic is showed in panel (b) for a pair of the tetrahedra across the boundary to highlight the two spin configurations. (c) MR in the region of the anomaly from 0.45 K to 1 K. The solid lines were measured with ***B//*[111], whereas the dashed lines were measured with ***B//*[1-10]. The curves are vertically shifted for clarity and comparison purposes.